\documentclass[aps,prl,twocolumn,groupedaddress,showpacs]{revtex4}
\usepackage{graphicx}

\begin{document}

\title{High Density Mesoscopic Atom Clouds in a Holographic Atom Trap}

\author{ J. Sebby-Strabley}
\author{R. T. R. Newell}\thanks{Current address: LANL}
\author{J. O. Day}
\author{E. Brekke}
\author{T. G. Walker}

\affiliation{Department of Physics, University of Wisconsin-Madison,
		Madison, Wisconsin 53706}

\begin{abstract}We  demonstrate the production of micron-sized high density
atom clouds of interest for mesoscopic quantum information processing.  
We evaporate atoms from  60 $\mu$K, $3\times10^{14}$ atoms/cm$^{3}$ samples contained in a highly anisotropic
optical lattice  formed by interfering diffracted beams from a holographic phase plate. After evaporating to 1 $\mu$K by lowering the
confining potential, in less than a second the atom density reduces to $8\times10^{13}$ cm$^{-3}$  at a phase space density
approaching unity.  Adiabatic recompression of the atoms then increases the density to levels in excess of $1\times10^{15}$
cm$^{-3}$.  The resulting clouds are typically  8 $\mu$m in the longest dimension.  Such samples are small enough to enable
mesoscopic quantum manipulation using Rydberg blockade and have the high densities required to investigate new collision phenomena.
\end{abstract}


\maketitle

Two of the most important themes in current studies of ultracold incoherent matter are studies of plasmas and Rydberg atoms at low
temperature and high density, and use of mesoscopic samples for topics in quantum information processing.  Even at the relatively
modest 10$^{12}$ cm$^{-3}$ densities of standard magneto-optical traps, a wide variety of new phenomena have been  observed by
exciting atoms near the ionization limit\cite{Simien04}
.  Similarly, the use of atomic
ensembles with their collectively enhanced light-atom interactions have led to new developments such as collective spin squeezing
\cite{Hald99}, quantum memory
\cite{vanderWal03}, and single-photon generation\cite{Chou04}.  With these contexts in mind we report in this Letter new techniques
for producing high density ($>10^{15}$ cm$^{-3}$) mesoscopic (5-10 $\mu$m) samples ideal for use in both types of experiments.

In the context of high densities, we note that the  recent predictions of novel ultralong range Rydberg molecular
states \cite{Greene00,Boisseau02,Farooqi03} require extremely high densities to attain significant production rates.  These 
``trilobite'' molecules  arise from the Fermi point-like interactions between Rydberg and ground-state atoms.  At the densities
reported here, the Fermi shifts of the Rydberg levels are on the order of 100 MHz.  Similarly, the production of cold plasmas and
Rydberg gases at these densities promises to reveal new phenomena not accesible in conventional laser traps which have 100-1000 times
lower densities.


Regarding mesoscopic atom samples,  Lukin {\it et al.}  
\cite{Lukin01} recently proposed manipulating quantum information  using the very long range dipole-dipole
interactions produced by Rydberg atoms. In brief, the excitation of a single Rydberg atom strongly suppresses the excitation of other
atoms within its range of influence.  If the size of a sample of atoms is less than the  blockade range $R$, then the
accessible quantum states of the ensemble are limited to states of zero or one atom excitations.  Using Rydberg states as
intermediate states in Rabi manipulation of two hyperfine ground states then allows the production of stable but highly entangled 
collective excitations of the ensembles.  For MHz rate quantum manipulations with cw lasers the value of $R$ required is  a few
$\mu$m at principal quantum numbers $n>50$ \cite{Walker04}.  The clouds produced in this experiment reach this length scale.

Recent experiments have demonstrated dramatic suppression of pulsed Rydberg state excitation in magneto-optical traps of
size
$L\gg R$
\cite{Tong04,Singer04}.  In these experiments, the number of excited Rydberg atoms saturates at fluence values much less than
expected for isolated atoms due to the Rydberg-Rydberg interactions shifting atoms out of resonance with the exciting lasers.  If the
Rydberg blockade is effective, the number of excited atoms should be limited to roughly $(L/R)^3$.  If the blockade is not
complete, saturation still occurs but with a greater number of excited atoms.  Thus to differentiate blockade and suppression
requires production of mesoscopic samples of size $L\sim R$ \cite{fosternote}.

In this Letter we present a method for producing high density elliptical mesoscopic atom samples with semi-major axes $\sigma$ on the
order of
$R$ using rapid evaporative cooling of $^{87}$Rb atoms from a  Holographic Atom Trap (HAT), followed by adiabatic recompression.  We
demonstrate densities in excess of 10$^{15}$ cm$^{-3}$, the highest cold atom densities attained for incoherent matter.   The
recompression stage can produce clouds of radius as small as $5.6$ $\mu$m, sufficiently small to be sensitive to single atom Rydberg
blockade.

\begin{figure}[htb]
\includegraphics[scale=0.25]{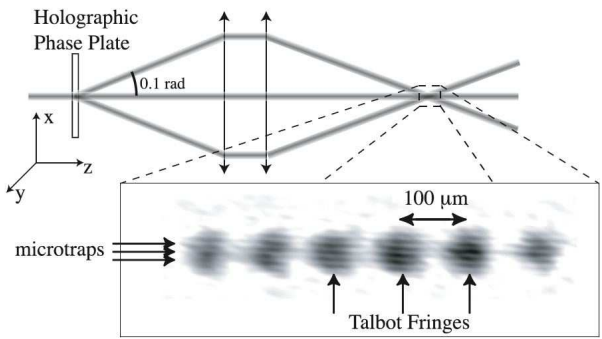}
\caption{Holographic Atom Trap: five laser beams diffracted from a phase plate are imaged onto a MOT cloud.  Atoms collect in
the intensity maxima of the interference pattern of the beams.}\label{hat}
\end{figure}

The HAT, described in detail
elsewhere
\cite{Newell03}, is a lattice of interference fringes produced by imaging 5 diffracted orders (zeroth order plus 4 equal intensity
first order beams) from a holographic phase plate. The laser used is an 18W cw flashlamp-pumped
Nd:YAG laser at 1.064
$\mu m$, intensity stabilized and controlled using an acousto-optic modulator feedback system.  Along the propagation axis  the
Talbot effect gives rise to a series of interference fringes.  Each Talbot fringe contains a lattice with a unit
cell of  10
$\mu$m $\times$ 10
$\mu$m $\times$ 100
$\mu$m as illustrated in Figure~\ref{hat}. The five diffracted orders are focussed to approximately 90 $\mu$m waists at the region of
intersection, giving a typical full-intensity trap depth of $U_0=600$ $\mu$K, and trap oscillation frequencies of $18.4\pm1.2$ kHz,
$18.4\pm1.2$ kHz, and $735\pm62$ kHz.  Atoms are loaded into the HAT from a forced-dark-spot
$F=1$ magneto-optical trap
\cite{Anderson94}.

We use absorption imaging to characterize the spatial distribution of the atoms in the HAT, to make absolute measurements of the
number of atoms, and to measure the atomic temperature via time-of-flight techniques.   The HAT is first turned off in 10 $\mu$s in
order to eliminate AC Stark shifts and excited state hyperfine mixing.  Then a 150 $\mu$s pulse of light from a diode laser tuned to
the 5S$_{1/2}(F=1)\rightarrow $ 5P$_{1/2}(F'=2)$ transition passes through the atoms which are imaged onto a CCD camera.  The imaging
lens system is a  pair of commercial achromats that give an aberration-limited resolution of approximately 5 $\mu$m. Images are 
analyzed to deduce the number and distribution of atoms contained in the different microtraps  We make use of a calibrated 
absorption method for absolute number measurements \cite{Gibble92}. When the fluence of the imaging pulse is sufficient to remove all
the atoms from the $F=1$ state, the average number of photons absorbed per atom is  given simply from the fluorescence branching
ratios to be 2.  The  number of atoms is then directly determined from the camera quantum efficiency and the
transmission of the lenses.  For small numbers of atoms and for time-of-flight temperature measurements, an additional laser tuned to
repump the atoms back to $F=1$ can be used to artificially increase the number of photons absorbed per atom.

\begin{figure}[htb]
\includegraphics[scale=0.28]{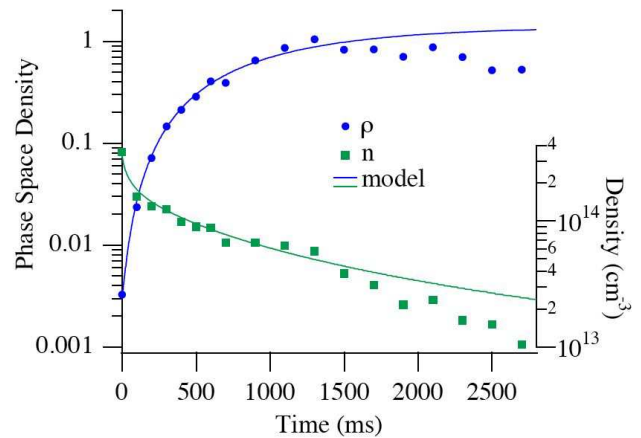}
\caption{During forced evaporation the  phase space density $\rho$ approaches unity while the density $n$ decreases  with
atom loss. }\label{evap}
\end{figure}

After loading atoms
into the HAT at densities of
$3\times10^{14}$ cm$^{-3}$ and temperatures of 60 $\mu$K, the atoms are distributed over typically five Talbot fringes, with each
Talbot fringe containing typically 25 occupied microtraps of slightly differing trap depths.
Atoms in these microtraps have  sufficiently high collision rates  (20,000/s) to initiate
forced evaporative cooling. Using established protocols \cite{Davis95,OHara01}, we gently reduce the HAT intensity, allowing high
kinetic energy atoms to escape while retaining low energy atoms and thereby increasing phase space density. The
evaporation process tightly couples the trap depth $U$ and the temperature so that $T\approx U/10$.  During evaporation, the density
$n$ scales as the atom number $N$; the phase space density $\rho$ scales as $ N \nu^3/T^3\propto n/U^{3/2}$. Even though the density
and
 the collision rates decrease with time, they are sufficiently high in the HAT that the limit on the evaporation speed is
the required adiabaticity of the z-motion and the desire not to remove too many atoms.      An interesting
feature of the HAT is the inequivalence of the various trapping sites:  sites towards the edge of the trap are more weakly bound than
those at the center.  As the evaporation proceeds  atom loss from the outer sites is greater, causing the fraction of atoms
contained in the central microtrap to increase from initially 6\% to nearly 15\%.  Atom  densities
and phase space densities as a function of time are shown in Figure~\ref{evap}. 

After evaporating to a final temperature $T_e$ and density $n_e$, typically $1$ $\mu$K and $8\times
10^{13}$ cm$^{-3}$, we recompress the cloud by adiabatically returning  the trap depth to its maximum value $U_0$.  Since the trap
depth increases more rapidly than the temperature, this shuts off evaporation, holding the number of atoms constant in the absence of
loss mechanisms. The phase space density  is conserved for an adiabatic process, thus $T\propto\nu\propto U^{1/2}$ and 
$n\propto N
\nu^3/T^{3/2}$.  As a result,  the recompressed density is
$n_r=n_e(U_0/U_e)^{3/4}$.  In our case, this gives a factor of 20 density increase, consistent with our measured final density of
$1.8\pm0.5\times 10^{15}$/cm$^3$.

\begin{figure}[htb]
\includegraphics[scale=0.28]{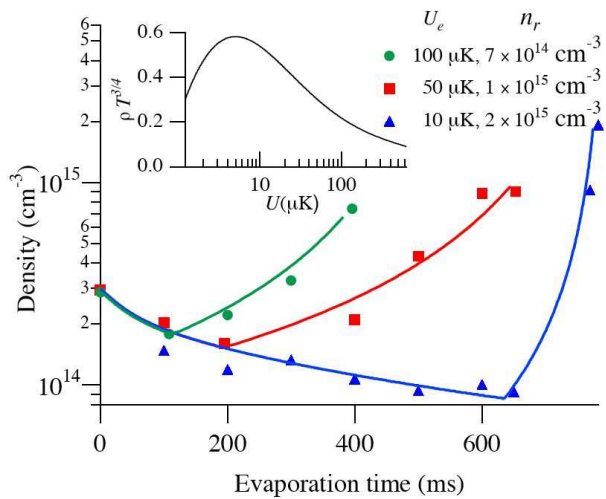}
\caption{ Compression data and model (solid line) for 3 values of $U_e$. The inset shows how the key parameter $\rho T^{3/4}$ varies
with trap depth.}\label{recompress}
\end{figure}

A simple argument shows that the final attainable density using this method depends not only on the phase space density $\rho_e$
achieved from evaporation, but the temperature $T_e$ as well.  The density $n_e$ after evaporation is proportional to
the product
$\rho_e T_e^{3/2}$.  The compression factor is $\left(U_0/U_e\right)^{3/4}\propto T_e^{-3/4}$ for fixed $U_0$, hence  the highest
density is achieved when evaporating to a $U_e$ which maximizes the function $\rho_e T_e^{3/4}$ (see inset to Fig.~\ref{recompress}). 
Continuing to lower
$U_e$ does not  achieve higher densities.  Evaporating to lower trap depths takes increasingly more time due to the adiabatic
constraints.  Losses  from heating mechanisms and background collisions  cause $\rho_e T_e^{3/4}$ to slowly decrease.  This
effectively determines the optimum value of $U_e$ at which to start the recompression.  Even with this constraint on $U_e$ it is
interesting to point out  the wide range of densities attainable  with this method.   

We show recompression data
 for various values of $U_e$ in Figure~\ref{recompress}. Like evaporation, the limiting timescale for
recompression is adiabaticity.  However, unlike evaporation, the number of atoms is conserved so the potential can be ramped up
rapidly.   Figure 3 shows that even at the lowest values of $U_e$  densities over $10^{15}$ cm$^{-3}$ can be achieved in less than 900
ms.  The adiabatic constraints on the $U_e=10$ $\mu$K data in Figure 3 were
purposely lightened to limit losses due to background collisions and heating mechanisms.  The effect is a small breakdown of the
scaling ratios which assume perfect adiabaticity.

\begin{figure}[htb]
\includegraphics[scale=0.28]{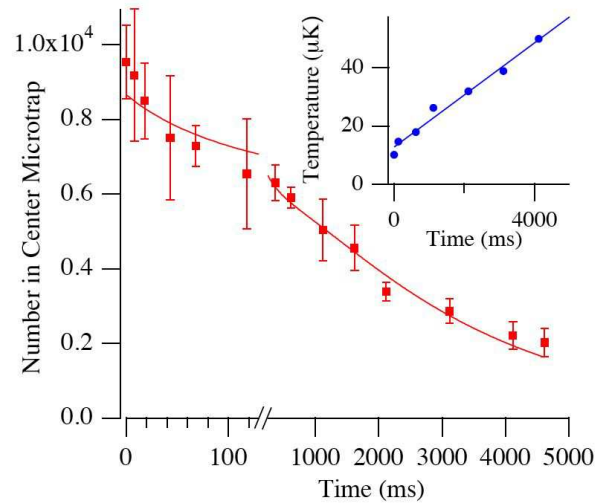}
\caption{After recompression we observe rapid atom loss due to 3-body recombination. Since there is no  evaporation from the
recompressed clouds we can measure trap heating rates by directly measuring dT/dt (inset).}\label{recombination}
\end{figure}

The high densities are deduced from measurements of the number of atoms, the fraction of atoms in each microtrap, the temperature, and
direct measurements of the trap spring constants.  The latter are obtained using the parametric heating method \cite{Friebel98}.  We
confirm the inferred densities using measurements of known 3-body recombination rates and by imaging the z-axis spatial
distribution.  The three-body recombination rates are determined by measuring the number of atoms in the recompressed central
microtrap as a function of time after recompression: $dN/dt=-K\int n(t)^3 dV-\Gamma N$, where $K$ is the 3-body recombination rate
coefficient and
$\Gamma$ the loss rate due to background collisions. Data are shown in Figure~\ref{recombination}.     After recompression
the temperature is too small to allow evaporation.  However, heating mechanisms  increase the temperature with time.  As a result,
$n(t)$ decreases  due to both recombination losses and temperature increases.  Taking these effects into account, we find
$K=3.5\pm1.9\times 10^{-29}$ cm$^6$/s, in close agreement with previous measurements
\cite{Burt97,Tolra04}.  We have measured the rate in a
magnetic field of 2.5 Gauss typically present in magnetic traps.  At this field we measure $K=4.8\pm2.3\times 10^{-29}$ cm$^6$/s. 
These results are extremely sensitive to density errors;  the agreement with previous experiments is
confirmation of the reliability of the density measurements.

We have also confirmed the densities by measuring  cloud sizes.  After evaporation and recompression, the semi-minor axis of the cloud
 is only 430 nm, too small to optically resolve.  However, the z-axis size of 10.8 $\mu$m at 13 $\mu$K is resolvable. 
The measured z-axis sizes agree to within 0.5 microns with the value $\sigma=\sqrt{2T/m\omega^2}$ expected from the frequency
and temperature measurements, once optical resolution and motion of the atoms during the imaging pulse are taken into account.

In addition to using evaporation followed by recompression to produce high densities, it can be used to produce strong confinement of
the atoms.  During evaporation, the spatial dimensions $\sigma$ of the atom clouds are unchanged since both $T$
and $\omega^2$ scale linearly with trap depth.  During recompression $T\propto\omega$ so that 
$\sigma\propto(U_0/U_b)^{1/4}$ yields up to a factor of  4 in size reduction for trap depth ratios of 300 that we have produced in our
system.  The smallest clouds we can directly observe are inferred from this ratio and the measured trap depth of 2 $\mu$K  to
be $\sigma=5.6$
$\mu$m.  The smallest clouds for which we have a direct temperature measurement have  $\sigma=8$ $\mu$m. 

We now evaluate these results in the context of Rydberg blockade.  We imagine excitation to $n$s Rydberg states with a Gaussian beam
of waist $10$ $\mu$m.  For non-uniform excitation, the appropriate figure of merit is the  Rydberg-Rydberg level shift
$\overline{\Omega}$ defined via $1/\overline{\Omega}^2=\langle \left|1/\Omega_{ij}\right|^2\rangle_{}$
averaged over all atom pairs $ij$.  Using calculated ns-ns potentials \cite{Walker04} with $n=(70,95)$, we find
$\overline{\Omega}=(2.0,62)$ MHz for
$\sigma=5.6 $ $\mu$m, and $\overline{\Omega}=(0.7,21)$ MHz
for
$\sigma=8.0 $ $\mu$m.  For a collective 2 $\mu$s  $\pi$-pulse, the probability of double excitation even in the worst case of
$\overline{\Omega}=0.7$ MHz is 6\%.  Thus these clouds are extremely well-suited to investigate high speed, high fidelity collective
coherent quantum manipulations.


For some applications, it may be useful to isolate a single microtrap from the others.  To do this, we exploit the anisotropy of the
microtraps to parametrically heat the atoms in the outer microtraps with little disturbance to the atoms contained in the central
microtrap.  The quality factor for our parametric heating experiment is about 10 for the x-direction,
sufficient to give significant differentiation of the 3\% difference in oscillation frequencies between the center and outer
microtraps.  Thus we drive the parametric resonance at its low frequency tail, selectively heating and ejecting the atoms from the
outer microtraps while causing minor heating for the center microtrap.  Doing this, we have been able to increase the center well
fraction from 15\% to above 40\%.

In addition to its utility for generating high densities and small clouds, adiabatic recompression also allows direct measurements of
trap heating rates.  Under conditions of evaporation,  heating  mechanisms do not actually increase the
temperature when evaporation is rapid enough to recool the atoms by atom loss.  Thus heating and loss mechanisms can be difficult to
disentangle under conditions of evaporation.  However, evaporation no longer functions after recompression, and heating rates can be
deduced directly from measured temperature increases with time.  Example data are shown in the inset to Figure~\ref{recombination}. 
The deduced heating rate
$dE/dt=3  dT/dt$ is  a factor of 1.4 larger than the predictions of Bali {\it et al.} \cite{Bali99} for quantum diffractive
heating.  Indeed, taking this measured heating rate into account in our evaporation model, combined with the heating due to multiple
scattering as predicted by Beijerinck
\cite{Beijerinck00}, we generate the solid curve in Fig.~\ref{evap} that accounts for our measured phase space density as a
function of time.


The experiments we have described represent a robust method for producing the kinds of high density ($>10^{15}$ cm$^{-3}$) and
small size clouds ($< 8$ $\mu$m) of interest for Rydberg atom studies both in the high density regime of ultracold plasmas and
molecular spectroscopy, and in the small size regime relevant to coherent single-atom manipulation for quantum information processing.
Many other proposed applications, including deterministic single atom and photon sources \cite{Saffman02} and fast quantum state
detection and transmission schemes \cite{Saffman04}, require these types of sources for their operation.

\begin{acknowledgments}This work was supported by the NSF and NASA.  We appreciate  discussions with M.
Saffman, C. Greene, B. Esry, J. Thomas and H. Beijerinck.\end{acknowledgments}

\newcommand{\noopsort}[1]{} \newcommand{\printfirst}[2]{#1}
  \newcommand{\singleletter}[1]{#1} \newcommand{\switchargs}[2]{#2#1}


\begin{thebibliography}{29}
\expandafter\ifx\csname natexlab\endcsname\relax\def\natexlab#1{#1}\fi
\expandafter\ifx\csname bibnamefont\endcsname\relax
  \def\bibnamefont#1{#1}\fi
\expandafter\ifx\csname bibfnamefont\endcsname\relax
  \def\bibfnamefont#1{#1}\fi
\expandafter\ifx\csname citenamefont\endcsname\relax
  \def\citenamefont#1{#1}\fi
\expandafter\ifx\csname url\endcsname\relax
  \def\url#1{\texttt{#1}}\fi
\expandafter\ifx\csname urlprefix\endcsname\relax\def\urlprefix{URL }\fi
\providecommand{\bibinfo}[2]{#2}
\providecommand{\eprint}[2][]{\url{#2}}

\bibitem[{\citenamefont{Simien et~al.}(2004)\citenamefont{Simien, Chen, Gupta,
  Laha, Martinez, Mickelson, Nagel, and Killian}}]{Simien04}
\bibinfo{author}{\bibfnamefont{C.}~\bibnamefont{Simien}}
  \bibinfo{author}{{\it et al.}},
  \bibinfo{journal}{Phys. Rev. Lett.} \textbf{\bibinfo{volume}{92}},
  \bibinfo{pages}{143001} (\bibinfo{year}{2004});
\bibinfo{author}{\bibfnamefont{T.}~\bibnamefont{Pohl}},
  \bibinfo{author}{\bibfnamefont{T.}~\bibnamefont{Pattard}}, \bibnamefont{and}
  \bibinfo{author}{\bibfnamefont{J. M.}~\bibnamefont{Rost}},
  \bibinfo{journal}{Phys. Rev. Lett.} \textbf{\bibinfo{volume}{92}},
  \bibinfo{pages}{155003} (\bibinfo{year}{2004});
\bibinfo{author}{\bibfnamefont{A.}~\bibnamefont{Walz-Flannigan}},
  \bibinfo{author}{\bibfnamefont{J.}~\bibnamefont{Guest}},
  \bibinfo{author}{\bibfnamefont{J.-H.} \bibnamefont{Choi}}, \bibnamefont{and}
  \bibinfo{author}{\bibfnamefont{G.}~\bibnamefont{Raithel}},
  \bibinfo{journal}{Phys. Rev. A} \textbf{\bibinfo{volume}{v 69}},
  \bibinfo{pages}{634051} (\bibinfo{year}{2004});
\bibinfo{author}{\bibfnamefont{T.}~\bibnamefont{Gallagher}}
  \bibinfo{author}{{\it et al.}},
  \bibinfo{journal}{J. Opt. Sci. Amer. B} \textbf{\bibinfo{volume}{20}},
  \bibinfo{pages}{1091} (\bibinfo{year}{2003});
\bibinfo{author}{\bibfnamefont{J.~L.} \bibnamefont{Roberts}},
  \bibinfo{author}{\bibfnamefont{C.~F.} \bibnamefont{Fertig}},
  \bibinfo{author}{\bibfnamefont{M.~L.} \bibnamefont{Lim}}, \bibnamefont{and}
  \bibinfo{author}{\bibfnamefont{S.~L.} \bibnamefont{Rolston}},
  \bibinfo{journal}{physics/040204}  (\bibinfo{year}{2004}).

\bibitem[{\citenamefont{Hald et~al.}(1999)\citenamefont{Hald, Sorensen, Schori,
  and Polzik}}]{Hald99}
\bibinfo{author}{\bibfnamefont{J.}~\bibnamefont{Hald}},
  \bibinfo{author}{\bibfnamefont{J. L.}~\bibnamefont{Sorensen}},
  \bibinfo{author}{\bibfnamefont{C.}~\bibnamefont{Schori}}, \bibnamefont{and}
  \bibinfo{author}{\bibfnamefont{E. S.}~\bibnamefont{Polzik}},
  \bibinfo{journal}{Phys. Rev. Lett.} \textbf{\bibinfo{volume}{83}},
  \bibinfo{pages}{1319} (\bibinfo{year}{1999});
\bibinfo{author}{\bibfnamefont{A.}~\bibnamefont{Kuzmich}},
  \bibinfo{author}{\bibfnamefont{L.}~\bibnamefont{Mandel}}, \bibnamefont{and}
  \bibinfo{author}{\bibfnamefont{N. P.}~\bibnamefont{Bigelow}},
  \bibinfo{journal}{Phys. Rev. Lett.} \textbf{\bibinfo{volume}{85}},
  \bibinfo{pages}{1594} (\bibinfo{year}{2000}).

\bibitem[{\citenamefont{van~der Wal et~al.}(2003)\citenamefont{van~der Wal,
  M.D., Andre, Walsworth, Phillips, Zibrov, and Lukin}}]{vanderWal03}
\bibinfo{author}{\bibfnamefont{C.~H.} \bibnamefont{van~der Wal}}
  \bibinfo{author}{{\it et al.}},
  \bibinfo{journal}{Science} \textbf{\bibinfo{volume}{301}},
  \bibinfo{pages}{196} (\bibinfo{year}{2003}).

\bibitem[{\citenamefont{Chou et~al.}(2004)\citenamefont{Chou, Polyakov,
  Kuzmich, and Kimble}}]{Chou04}
\bibinfo{author}{\bibfnamefont{C.~W.} \bibnamefont{Chou}},
  \bibinfo{author}{\bibfnamefont{S.~V.} \bibnamefont{Polyakov}},
  \bibinfo{author}{\bibfnamefont{A.}~\bibnamefont{Kuzmich}}, \bibnamefont{and}
  \bibinfo{author}{\bibfnamefont{H.~J.} \bibnamefont{Kimble}},
  \bibinfo{journal}{Phys. Rev. Lett.} \textbf{\bibinfo{volume}{92}},
  \bibinfo{pages}{213601} (\bibinfo{year}{2004}).

\bibitem[{\citenamefont{Greene et~al.}(2000)\citenamefont{Greene, Dickinson,
  and Sadeghpour}}]{Greene00}
\bibinfo{author}{\bibfnamefont{C.~H.} \bibnamefont{Greene}},
  \bibinfo{author}{\bibfnamefont{A.~S.} \bibnamefont{Dickinson}},
  \bibnamefont{and} \bibinfo{author}{\bibfnamefont{H.~R.}
  \bibnamefont{Sadeghpour}}, \bibinfo{journal}{Phys. Rev. Lett.}
  \textbf{\bibinfo{volume}{85}}, \bibinfo{pages}{2458} (\bibinfo{year}{2000}).

\bibitem[{\citenamefont{Boisseau et~al.}(2002)\citenamefont{Boisseau, Simbotin,
  and Cote}}]{Boisseau02}
\bibinfo{author}{\bibfnamefont{C.}~\bibnamefont{Boisseau}},
  \bibinfo{author}{\bibfnamefont{I.}~\bibnamefont{Simbotin}}, \bibnamefont{and}
  \bibinfo{author}{\bibfnamefont{R.}~\bibnamefont{Cote}},
  \bibinfo{journal}{Phys. Rev. Lett.} \textbf{\bibinfo{volume}{88}},
  \bibinfo{pages}{133004} (\bibinfo{year}{2002}).

\bibitem[{\citenamefont{Farooqi et~al.}(2003)\citenamefont{Farooqi, Tong,
  Krishnan, Stanojevic, Zhang, Ensher, Estrin, Boisseau, Cote, Eyler
  et~al.}}]{Farooqi03}
\bibinfo{author}{\bibfnamefont{S.}~\bibnamefont{Farooqi}}
  \bibinfo{author}{{\it et al.}}, \bibinfo{journal}{Phys. Rev. Lett.}
  \textbf{\bibinfo{volume}{91}}, \bibinfo{pages}{183002}
  (\bibinfo{year}{2003}).

\bibitem[{\citenamefont{Lukin et~al.}(2001)\citenamefont{Lukin, Fleischhauer,
  Cote, Duan, Jaksch, Cirac, and Zoller}}]{Lukin01}
\bibinfo{author}{\bibfnamefont{M.~D.} \bibnamefont{Lukin}}
  \bibinfo{author}{{\it et al.}},
  \bibinfo{journal}{Phys. Rev. Lett.} \textbf{\bibinfo{volume}{87}},
  \bibinfo{pages}{37901} (\bibinfo{year}{2001}).

\bibitem[{\citenamefont{Walker and Saffman}(2004)}]{Walker04}
\bibinfo{author}{\bibfnamefont{T.~G.} \bibnamefont{Walker}} \bibnamefont{and}
  \bibinfo{author}{\bibfnamefont{M.}~\bibnamefont{Saffman}},
  \bibinfo{journal}{physics/0407048}  (\bibinfo{year}{2004}).

\bibitem[{\citenamefont{Tong et~al.}(2004)\citenamefont{Tong, Farooqi,
  Stanojevic, Krishnan, Zhang, Cote, Eyler, and Gould}}]{Tong04}
\bibinfo{author}{\bibfnamefont{D.}~\bibnamefont{Tong}}
  \bibinfo{author}{{\it et al.}},
  \bibinfo{journal}{physics/0402113}  (\bibinfo{year}{2004}).

\bibitem[{\citenamefont{Singer et~al.}(2004)\citenamefont{Singer, Reetz-Lamour,
  Amthor, Marcassa, and Weidemuller}}]{Singer04}
\bibinfo{author}{\bibfnamefont{K.}~\bibnamefont{Singer}}
  \bibinfo{author}{{\it et al.}},
  \bibinfo{journal}{physics/0402113}  (\bibinfo{year}{2004}).

\bibitem[{fos()}]{fosternote}
\bibinfo{note}{An additional requirement is that the Rydberg-Rydberg
  interaction has no dipole-dipole zeros. See Ref.~\protect\cite{Walker04}.}

\bibitem[{\citenamefont{Newell et~al.}(2003)\citenamefont{Newell, Sebby, and
  Walker}}]{Newell03}
\bibinfo{author}{\bibfnamefont{R.}~\bibnamefont{Newell}},
  \bibinfo{author}{\bibfnamefont{J.}~\bibnamefont{Sebby}}, \bibnamefont{and}
  \bibinfo{author}{\bibfnamefont{T.}~\bibnamefont{Walker}},
  \bibinfo{journal}{Opt. Lett.} \textbf{\bibinfo{volume}{28}},
  \bibinfo{pages}{1266} (\bibinfo{year}{2003}).

\bibitem[{\citenamefont{Anderson et~al.}(1994)\citenamefont{Anderson, Petrich,
  Ensher, and Cornell}}]{Anderson94}
\bibinfo{author}{\bibfnamefont{M.~H.} \bibnamefont{Anderson}},
  \bibinfo{author}{\bibfnamefont{W.}~\bibnamefont{Petrich}},
  \bibinfo{author}{\bibfnamefont{J.~R.} \bibnamefont{Ensher}},
  \bibnamefont{and} \bibinfo{author}{\bibfnamefont{E.~A.}
  \bibnamefont{Cornell}}, \bibinfo{journal}{Phys. Rev. A}
  \textbf{\bibinfo{volume}{50}}, \bibinfo{pages}{R3597} (\bibinfo{year}{1994}).

\bibitem[{\citenamefont{Gibble et~al.}(1992)\citenamefont{Gibble, Kasapi, and
  Chu}}]{Gibble92}
\bibinfo{author}{\bibfnamefont{K.~E.} \bibnamefont{Gibble}},
  \bibinfo{author}{\bibfnamefont{S.}~\bibnamefont{Kasapi}}, \bibnamefont{and}
  \bibinfo{author}{\bibfnamefont{S.}~\bibnamefont{Chu}}, \bibinfo{journal}{Opt.
  Lett.} \textbf{\bibinfo{volume}{17}}, \bibinfo{pages}{526}
  (\bibinfo{year}{1992}).

\bibitem[{\citenamefont{Davis et~al.}(1995)\citenamefont{Davis, Mewes, and
  Ketterle}}]{Davis95}
\bibinfo{author}{\bibfnamefont{K.}~\bibnamefont{Davis}},
  \bibinfo{author}{\bibfnamefont{M.-O.} \bibnamefont{Mewes}}, \bibnamefont{and}
  \bibinfo{author}{\bibfnamefont{W.}~\bibnamefont{Ketterle}},
  \bibinfo{journal}{Appl. Phys. B} \textbf{\bibinfo{volume}{60}},
  \bibinfo{pages}{155} (\bibinfo{year}{1995}).

\bibitem[{\citenamefont{O'Hara et~al.}(2001)\citenamefont{O'Hara, Gehm,
  Granade, and Thomas}}]{OHara01}
\bibinfo{author}{\bibfnamefont{K.~M.} \bibnamefont{O'Hara}},
  \bibinfo{author}{\bibfnamefont{M.~E.} \bibnamefont{Gehm}},
  \bibinfo{author}{\bibfnamefont{S.~R.} \bibnamefont{Granade}},
  \bibnamefont{and} \bibinfo{author}{\bibfnamefont{J.~E.}
  \bibnamefont{Thomas}}, \bibinfo{journal}{Phys. Rev. A}
  \textbf{\bibinfo{volume}{64}}, \bibinfo{pages}{51403} (\bibinfo{year}{2001}).

\bibitem[{\citenamefont{Friebel et~al.}(1998)\citenamefont{Friebel,
  Scheunemann, Walz, Hansch, and Weitz}}]{Friebel98}
\bibinfo{author}{\bibfnamefont{S.}~\bibnamefont{Friebel}}
  \bibinfo{author}{{\it et al.}},
  \bibinfo{journal}{Appl. Phys. B} \textbf{\bibinfo{volume}{B67}},
  \bibinfo{pages}{699} (\bibinfo{year}{1998}).

\bibitem[{\citenamefont{Burt et~al.}(1997)\citenamefont{Burt, Ghrist, Myatt,
  Holland, Cornell, and Wieman}}]{Burt97}
\bibinfo{author}{\bibfnamefont{E.~A.} \bibnamefont{Burt}}
  \bibinfo{author}{{\it et al.}},
  \bibinfo{journal}{Phys. Rev. Lett.} \textbf{\bibinfo{volume}{79}},
  \bibinfo{pages}{337} (\bibinfo{year}{1997}).

\bibitem[{\citenamefont{Tolra et~al.}(2004)\citenamefont{Tolra, Ohara, Huckans,
  Phillips, Rolston, and Porto}}]{Tolra04}
\bibinfo{author}{\bibfnamefont{B.~L.} \bibnamefont{Tolra}}
  \bibinfo{author}{{\it et al.}},
  \bibinfo{journal}{Phys. Rev. Lett.} \textbf{\bibinfo{volume}{92}},
  \bibinfo{pages}{190401} (\bibinfo{year}{2004}).

\bibitem[{\citenamefont{Bali et~al.}(1999)\citenamefont{Bali, O'Hara, Gehm,
  Granade, and Thomas}}]{Bali99}
\bibinfo{author}{\bibfnamefont{S.}~\bibnamefont{Bali}}
  \bibinfo{author}{{\it et al.}},
 \bibinfo{journal}{Phys. Rev. A}
  \textbf{\bibinfo{volume}{60}}, \bibinfo{pages}{R29} (\bibinfo{year}{1999}).

\bibitem[{\citenamefont{Beijerinck}(2000)}]{Beijerinck00}
\bibinfo{author}{\bibfnamefont{H.~C.~W.} \bibnamefont{Beijerinck}},
  \bibinfo{journal}{Phys. Rev. A} \textbf{\bibinfo{volume}{62}},
  \bibinfo{pages}{63614} (\bibinfo{year}{2000}).

\bibitem[{\citenamefont{Saffman and Walker}(2002)}]{Saffman02}
\bibinfo{author}{\bibfnamefont{M.}~\bibnamefont{Saffman}} \bibnamefont{and}
  \bibinfo{author}{\bibfnamefont{T.~G.} \bibnamefont{Walker}},
  \bibinfo{journal}{Phys. Rev. A} \textbf{\bibinfo{volume}{66}},
  \bibinfo{pages}{65403} (\bibinfo{year}{2002}).

\bibitem[{\citenamefont{Saffman and Walker}(2004)}]{Saffman04}
\bibinfo{author}{\bibfnamefont{M.}~\bibnamefont{Saffman}} \bibnamefont{and}
  \bibinfo{author}{\bibfnamefont{T.~G.} \bibnamefont{Walker}},
  \bibinfo{journal}{quant-ph/0402111}  (\bibinfo{year}{2004}).

\end{thebibliography}
\end{document}